\documentstyle[multicol,prl,aps,epsfig]{revtex}
\begin{document}

\def\gapprox{\lower.4ex\hbox{$\;\buildrel >\over{\scriptstyle\sim}\;$}}
\def\lapprox{\lower.4ex\hbox{$\;\buildrel <\over{\scriptstyle\sim}\;$}}
\def\bB{{\bf B}}
\def\bE{{\bf E}}
\def\bV{{\bf V}}
\def\cE{{E}}
\def\adjust{{\vspace*{-0.1in}}}

\title{Plasmoid impacts on neutron stars and highest energy cosmic rays}
\author{C.\ Litwin and R.\ Rosner}
\address{Department of Astronomy \& Astrophysics, The University of
Chicago, 5640 South Ellis Avenue, Chicago IL 60637}

\date{To be published in Physical Review Letters}
%\date{1 April 2001}
\maketitle

\begin{abstract}
%\vspace*{-0.1in}
\widetext Particle acceleration by electrostatic polarization fields
that arise in plasmas streaming across magnetic fields is discussed as
a possible acceleration mechanism of highest-energy ($\gapprox
10^{20}$ eV) cosmic rays.  Specifically, plasmoids arising in
planetoid impacts onto neutron star magnetospheres are considered.  We
find that such impacts at plausible rates may account for the observed
flux and energy spectrum of the highest energy cosmic
rays.\vspace*{0.in} \vspace*{-18pt}\end{abstract}

%\mediumtext

%]
%% 
 %%

 %%

\begin{multicols}{2}
%\raggedcolumns
\narrowtext

The origin of ultra-high energy cosmic rays (UHECRs), with energies up
to and exceeding $10^{20}$ eV \cite{efim91,haya94,bird95}, remains unknown:
the commonly invoked diffusive first-order Fermi acceleration of
cosmic rays in a supernova shock \cite{axfo77,krym77,bell78,blan78},
can accelerate particles to at most $\sim 10^{15}-10^{16}$ eV
\cite{laga83}.  While additional acceleration to energies
$\sim 100$ times higher by the electric field in a pulsar-driven
supernova remnant has been proposed \cite{bell92}, these energies are
still much below the highest observed energies.  Other models invoke
Fermi acceleration associated with cosmological gamma ray burst
sources \cite{waxman} and a decay of supermassive X particles of grand
unified field theories \cite{sigl95}.  In this Letter we consider a
different acceleration mechanism, based on charge
polarization arising in plasmoids impacting neutron star
magnetospheres.

It is well known \cite{chandra60,schmidt60} that an electrostatic
field arises in bounded plasmas moving across the magnetic field at
sub-Alfv\'enic velocities.  The reason for this is plasma polarization
caused by opposing gravitational and polarization drifts of electrons
and ions that lead to the appearance of net charge near the plasma
boundary.  If the plasma density $\rho$ is so high that the transverse
susceptibility $\chi_{\perp}\equiv 4\pi\rho c^{2}/B^{2}\gg 1$, then
the electrostatic field $-\nabla\Phi =-\bV\times\bB /c$ where $\bV$ is
plasma flow velocity and $\bB$ is the magnetic field; the potential
drop across the plasmoid of width $h$ in the cross-field direction
(denoted by ${\perp}$) is $2\Phi_{0} = hV_{\perp}B/c$ (Fig.  1).

Outside the plasmoid, the stray electrostatic field has a large
component parallel to the magnetic field which causes particle
acceleration along the field lines.  This phenomenon has been observed
in numerical simulations \cite{galv91,neub92,kita96} which showed that
charge layers can accelerate particles to relativistic energies even
for relatively slow (sub-Alfv\'enic) plasma flows; the accelerated
particle energy $\cE$ is $\sim q\Phi_{0}$, where $q$ is the particle
charge.  This estimate for $\cE$ derives from the fact that the
electrostatic field is dipole-like outside the plasma, giving rise to
the potential drop on the order of $\Phi_{0}$ along B \cite{neub92}.

The process of particle acceleration is transient: The energetic 
particle outflow from boundary layers of the plasmoid gives rise to 
plasma current; the resulting force decelerates the plasmoid 
cross-field motion (see below).

As an example, consider a plasmoid with $h\sim 10$ km infalling at the
the free-fall velocity onto the surface of a canonical neutron star of mass
$M_*=1.4M_\odot$, radius $R_*=$ 10 km and surface magnetic field
$B_*=5\times 10^{12}$ G \cite{shap83}.  The accelerating potential
$\Phi_{0}\sim 10^{21}$ V is then sufficient to accelerate protons to
UHECR energies, and heavier nuclei to even higher energies.

\begin{figure}
\epsfig{file=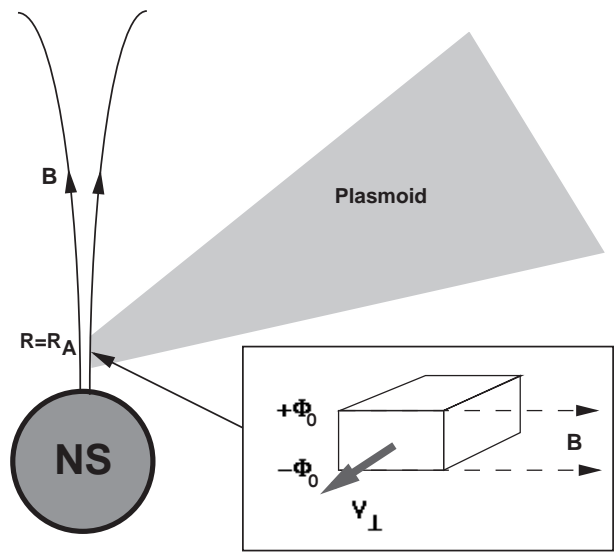,width=8.5cm} \caption{Schematic of a plasmoid 
infall on the neutron star.
\label{fig1}}\end{figure}

Several questions arise immediately: Can the infalling plasmoid have
the required width?  Can accelerated particles escape the neutron star
magnetic field?  And can the above-described mechanism give rise to
the observed energetic particle flux and spectrum?  In the remainder
of this Letter we address these issues.

Because of their large Larmor radii, the neutron star magnetic field
(assumed to be dipolar) would not confine electrons or protons in the
considered energy range (although it would confine heavier nuclei). 
Nevertheless, only a small fraction can escape, due to large radiative
losses: the curvature/synchrotron radiation results in a slowing down
length that is small when compared to the size of the magnetosphere. 
Only in regions where the magnetic field curvature is small, viz.,
near the magnetic axis, are radiative losses not prohibitive.

In order to make our discussion more quantitative, let us consider a
specific model.  Suppose an iron ($Z=56$) planetoid (or a
planetesimal), with the characteristic mass $M_{a}\sim
10^{22}-10^{24}$ g, such as those previously discussed in different
contexts \cite{colgate81,lin91,katz94,wasserman94,colgate96}, impacts
at the free-fall velocity onto an isolated, slowly rotating neutron star. 
We assume that the impact is grazing and occurs near the magnetic pole
at a large angle $\psi$ ($\gapprox\pi/6$) to the magnetic field (see Fig. 
1).

During the infall the planetoid becomes fragmented and compressed by
tidal forces, and ionized, by increased temperature
and large motional electric field (the Stark shift becomes
comparable to the ionization energy at a distance $R\sim 20-100
R_{\ast}$ from the neutron star center).  At distances larger than the
Alfv\'en radius $R_{A}$ (i.e., the distance where the ram
pressure equals the magnetic pressure and the free-fall velocity
equals the Alfv\'en velocity), the plasmoid motion is expected to be
ballistic with the external magnetic field screened from the plasma
interior by surface currents \cite{lamb73}.  At $R\sim R_A$ the
external magnetic field is commonly believed to penetrate the plasma
(e.g., \cite{lamb73,burnard83,hameury86}).  The exact mechanism of
this penetration (which is likely to involve anomalous resistivity,
cf.  \cite{mishin86}) is not entirely understood.  Nevertheless, we
assume that at $R\approx R_{A}$ the plasma becomes ``threaded'' by the
magnetic field, polarizes and $E\times B$ drifts as described by
Chandrasekhar \cite{chandra60} and Schmidt \cite{schmidt60}.

We adopt the model of Colgate \& Petschek \cite{colgate81} to describe
the planetoid motion at large distances.  Because of a small impact
parameter, the planetoid infall is nearly radial.  The planetoid
becomes fragmented at the distance
$R_{b}=(\rho_{0}r_{0}^{2}GM_*/s)^{(1/3}$, where $\rho_{0}$ is the
density, $r_{0}$ is the radius and $s$ is the tensile strength of the
planetoid (for an Fe planetoid, $\rho_{0}=8$ g cm$^{-3}$ and $s\sim
10^{10}$ dyne cm$^{-2}$).  The planetoid material initially undergoes
a phase of incompressible elongation; it then becomes elongated and
compressed at $R<R_{i}=\kappa R_{b}$ where $\kappa
=(5s/8P_{0})^{{2/5}}$, with $P_{0}$ ($\sim 100 s$) being the
compressive strength.  For $R\ll R_{i}$, $\rho\approx
\rho_{0} (R_{i}/R)^{1/2}/4$.  Then 

\begin{eqnarray}\label{alfven}
\frac{R_{A}}{R_{\ast}}=\left(\frac{B_{\ast}^{2}}{\pi\rho_{0}V_{\ast}^{2}}\right)^{\frac{2}{9}}
\left(\frac{r_{0}}{R_{\ast}}\right)^{-\frac{2}{27}}
\left(\frac{2s}{\rho_{0}V_{\ast}^{2}\kappa^{3}}\right)^{\frac{1}{27}}\\
\frac{r_{A}}{R_{\ast}}=\sqrt{5}\left(\frac{B_{\ast}^{2}}{\pi\rho_{0}V_{\ast}^{2}}\right)^{\frac{1}{9}}
\left(\frac{r_{0}}{R_{\ast}}\right)^{\frac{17}{27}}
\left(\frac{2s}{\rho_{0}V_{\ast}^{2}\kappa^{33/20}}\right)^{\frac{5}{27}}\label{radius}
\end{eqnarray}

\noindent where ${r_{A}}$ is the planetoid radius at $R=R_{A}$.  We
estimate the electrostatic potential drop $\Phi_{A}
=\frac{1}{c}r_{A}V_{A}B_{A}$ where
$V_{A}=V_{\ast}(R_{\ast}/R_{A})^{1/2}$ and
$B_{A}=B_{\ast}(R_{\ast}/R_{A})^{3}$.

We require that $R_{A}>R_{\ast}$, which implies that
$$
B_{\ast}>B_{\rm min}=
\left(\frac{r_{0}}{R_{\ast}}\right)^{1/6}
\left(\frac{\rho_{0}V_{\ast}^{2}\kappa^{3}}{2s}\right)^{1/12}
\left({\pi\rho_{0}V_{\ast}^{2}}\right)^{1/2}
$$
For an Fe planetoid of mass $\sim 10^{23}$ g and $M_*=1.4M_\odot$,
$B_{\rm min}\approx 10^{12}$ G; we therefore limit our attention to
$B_{\ast}\sim 10^{12}-10^{14}$ G. For such magnetic fields, $r_{A}\sim
3-8$ km, $R_{A}\sim 10-90$ km and the susceptibility
$\chi_{\perp}(R_{A})\sim 3-30$.

In order to evaluate the effect of radiative losses on the energy of
particles escaping the neutron star magnetosphere we solve the
equation of motion of a charged particle including the radiation
reaction force:

\begin{figure}\adjust
\epsfig{file=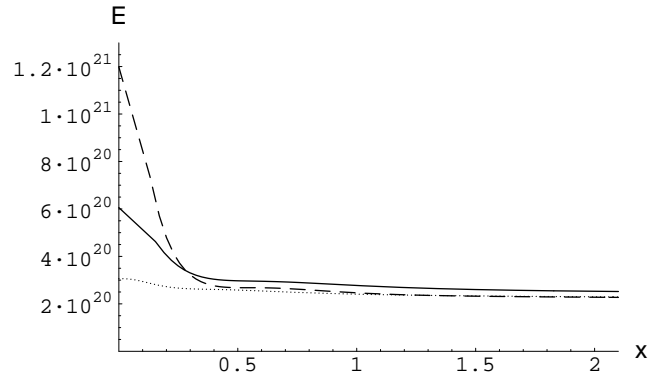,width=8.5cm} \caption{Particle energy as the 
function of distance $x=(R-R_{0})/R_{0}$ for $B_{\ast}=5\times 
10^{12}$ G, $\theta = 10^{-4}$ and $M_{a} =10^{22}$ g (dotted line), 
$M_{a} =10^{23}$ g (solid line) and $M_{a} =10^{24}$ g (dashed line).
\label{fig2}}\end{figure}

\begin{equation}
\frac{d\bf u}{d\tau}={\bf u}\times{\bf\Omega}+{\bf F}
\end{equation}

\noindent where $\tau$ is the proper time, ${\bf u}$ is the spatial
component of the four-velocity, ${\bf\Omega}=q\bB/mc$ is the
(vector) cyclotron frequency, $m$ is the particle mass and ${\bf F}$
is the radiation reaction force \cite{rohrlich65}; in the
ultra-relativistic limit considered here, ${\bf F}=-\lambda
({\bf\Omega}\times{\bf u})^{2}{\bf u}$ with $\lambda =2q^{2}/3mc^{5}$. 

We solve the above equation for a dipole magnetic field subject to the
initial value conditions corresponding to the motion of a particle
injected at $t=0$ in parallel with the magnetic field, with energy
$\cE_{0}=Ze\Phi_{A}$, at the radial distance $R_{0}=R_{A}$ and at an
angle $\theta$ with respect to the magnetic axis.  As we see in Fig. 
2, a particle is decelerated by the radiation reaction and emerges
from the magnetosphere with a fraction of its injection energy; the
emerging particle energy $\cE$ depends weakly on the planetoid mass. 
On the other hand, $\cE$ is sensitive to the injection angle $\theta$
(see Fig.  3).  For $B_{\ast}\sim 10^{12}-10^{14}$ G, $E>10^{19}$ eV
for angles less than $2-6\times 10^{-3}$.

We can now compute the energy spectrum of particles emerging from the
magnetosphere for a {\it single} impact event.  The {\it average}
number of particles per planetoid impact on the neutron star
(including impacts not in the vicinity of the magnetic axis), emitted
within solid angle $d\Omega$ about the magnetic axis is
$dN=N_{0}d\Omega /2\pi$, where $N_{0}$ is the total number of
accelerated particles per event (see below); we have exploited here
the fact that the planetoid angular extent is much larger than that of
the region in which particles with energies of interest ($\gapprox
10^{19}$ eV) are generated.  Thus the differential energy spectrum is
given by $dN/d\cE =N_{0}\sin\theta (\cE )d\theta/d\cE$.

In Fig.  4 we show the differential spectrum for magnetic fields
$B_{\ast}\sim 10^{12}-10^{14}$ G in the energy range
$10^{19}-10^{20.5}$ eV observed with AGASA. The spectrum can be well
approximated by the power law $dN/d{\cE}\sim\cE^{-\nu}$, with $\nu
=3.03$, $2.95$ and $2.89$ for $B=10^{12}$, $10^{13}$ and $10^{14}$ G,
respectively (the cut-off at $E=2\times 10^{20}$ eV for $B=10^{14}$ G
corresponds to $\theta =0$, cf.  Fig.  3).  Within the error bars,
this agrees with the power spectrum observed with AGASA
\cite{takeda98} ($\nu = 2.78^{+0.25}_{-0.33}$) and in the Akeno
experiment \cite{nagano92} ($\nu = 2.8\pm 0.3$).  Note, however, that
the spectrum of particles emerging from the neutron star magnetosphere
might differ from the spectrum observed on Earth, depending on
the UHECR confinement characteristics in the galactic magnetic
field (see later).

\begin{figure}\adjust
\epsfig{file=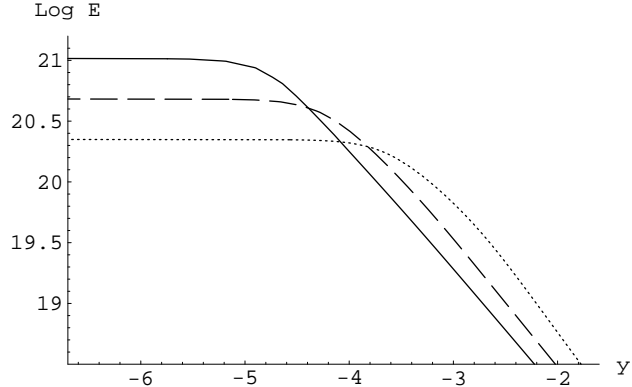,width=8.5cm}
\caption{Energy $E$ (in eV) of a particle emerging from the
magnetosphere as the function of the injection angle ($y\equiv\log\theta$),
for $B_{\ast}=10^{12}$ G (solid line), $B_{\ast}=10^{13}$ G (dashed
line) and $B_{\ast}=10^{14}$ G (dotted line).
\label{fig3}}\end{figure}

Let us now consider whether the mechanism discussed in the present
Letter can give rise to the observed flux of the highest energy cosmic
rays.  AGASA observations \cite{takeda98} imply that the number
density of particles with energies exceeding $10^{19}$ eV is
$n_{obs}\approx 6\times 10^{-29}$ cm$^{-3}$.  In order to determine
the density of UHECRs predicted by our model let us first find the
number of particles with energies greater than $E$ produced in
each impact event,
$$
N(E)=\int^{\infty}_{E}dE'\frac{dN}{dE'}=N_{0}[1-\cos\theta (E)]
$$

In order to find $N_{0}$ we first note that for the considered range
of parameters, the plasmoid is decelerated by the ${\bf J}\times\bB$
force before reaching the neutron star surface.  This can be seen as
follows: the field-aligned current carried by the energetic particle
outflow from boundary layers of the plasmoid has surface density
$\sigma c$ where surface charge density $\sigma = V_{\perp}B/4\pi c$. 
Because of charge conservation, this current equals the plasma
current, so that the plasma current density $J=2\sigma c/w$, where $w$
is the plasmoid width along the magnetic field.  Because of the ${\bf
J}\times\bB$ force the plasma is decelerated on the time scale $\tau_d
= \rho V_{\perp}c/JB=2\pi\rho c w/B^{2}$.  In the vicinity of
$R=R_{A}$, $\tau_d=cr_{A}/V_{A}^{2}$.  For characteristic fields and
planetoid masses considered in this Letter, the deceleration time is
shorter than the free-fall time; thus the cross-field motion of the
plasmoid is halted before it impacts the neutron star surface. 
Integrating the momentum balance equation (with gravitational and
pressure forces neglected as small compared to the ${\bf J}\times\bB$
force at $R<R_{A}$)

$$
\rho\frac{d\bf V}{dt}=\frac{1}{c}{\bf J}\times\bB
$$
over time and plasma volume one then finds that
$$
W_{A}=Q_{0}\Phi_{A}
$$
where $W_{A}=M_{a}V_{A}^{2}/2$ is the planetoid kinetic energy at $R=R_{A}$
and $Q_{0}=ZeN_{0}$ is the total charge carried by accelerated ions. 
Integrating the spectrum (Fig. 3), we find that $N(E=10^{19}
{\rm\ eV})\sim 0.2-1.4\times 10^{29}$ for $B_{\ast}\sim 10^{12}-10^{14}$ G.

\begin{figure}\adjust
\epsfig{file=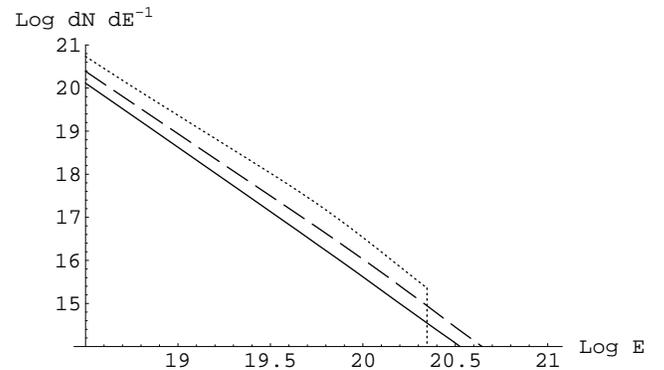,width=8.5cm} 
\caption{Differential energy spectrum for $B_{\ast}=10^{12}$ G (solid line), 
$B_{\ast}=10^{13}$ G (dashed line) and $B_{\ast}=10^{14}$ G (dotted line).
\label{fig4}}\end{figure}

In order to estimate the flux of UHECRs observed at Earth, we need to
know the number of neutron stars, the rate of impact events, and the 
confinement time of cosmic rays in the Galaxy; none of these are known
with any degree of certainty.

For the characteristic strength of the galactic magnetic field ($\sim$
3 $\mu$G) the Larmor radius of Fe nuclei with $E\lapprox 10^{20}$ eV
is $\lapprox 1$ kpc; thus such particles are confined by the galactic
magnetic field.  The density of UHECRs with energies greater than $E$
is therefore $n=\alpha\tau_{c}N(E)n_{NS}$, where $n_{_{NS}}$ is the
number density of neutron stars, $\alpha$ is the impact rate, and
$\tau_{c}$ is the confinement time of UHECRs in the galactic magnetic
field.  The density of neutron stars is estimated \cite{shap83} to be
$n_{_{NS}}\sim 2\cdot 10^{-3}$ pc$^{-3}$.  Thus $n=n_{\rm obs}$
requires $\alpha\tau_{c}\sim 6-40$.

The upper bound on the confinement time is given by the UHECR decay
time ($\sim 10^{16}$ s) due to photodisintegration by the infrared
background radiation \cite{puget76,stecker98}; the lower bound is set
by cross-field particle drifts due to the galactic magnetic field
inhomogeneity and curvature.  For example, a $10^{19}$ eV Fe nucleus
in a magnetic field with curvature $R_{c}\sim 10$ kpc would drift a
distance on the order of $R_{c}$ in 10$^{13}$ s; however, if the
magnetic field is twisted, the confinement time could be significantly
longer \cite{lau93}.  

The confinement time variation with particle energy determines the
difference between the spectrum at the source and the spectrum
observed on Earth and depends on the character of the global galactic
field as well as on the spectrum of magnetic fluctuations.  At lower
energies (below the ``knee'') the diffusion time is believed to be a
decreasing function of energy \cite{berezinskii90} which leads to the
observed spectrum that is steeper than the source spectrum.  For UHECR
energies ($>10^{19}$ eV), however, unlike at lower energies, the
gyroradius $\rho_{c}$ of an Fe nucleus is larger than the integral
length scale $L_{c}$ ($\sim$ 100 pc) of magnetic fluctuations in the
Galaxy \cite{parker79} which may lead to different confinement
characteristics \cite{confinement}.

Assuming now the confinement time $\tau_{c}\sim 10^{13}-10^{16}$ s,
we find the required impact rate to be one in $10^{6\pm 2}$ years per
neutron star (higher if the density of strongly magnetized neutron
stars is significantly lower than $n_{_{NS}}$).

It is rather difficult to assess whether this impact event rate is
plausible.  The rate of solid object impacts on neutron
stars has been a subject of widely varying estimates in the past
\cite{scha85,trem85,katz86,pine89,katz94,colgate96} in a different
context.  We do not attempt to make yet another estimate in the
present Letter and only note that the rate required by our model is
consistent with some of these previous estimates.

In summary, we have explored the plausibility of the acceleration by
the polarization electric field, which arises in plasma resulting from
planetoid accretion onto magnetized neutron stars, as the generation
mechanism for the cosmic rays with highest observed energies.  We
found that the source spectrum of particles generated by this
mechanism is similar to the observed spectrum; whether the resulting
spectrum observed on Earth will retain the same character is, at
present, an open question.  The calculated particle flux magnitude is
plausible, albeit quite uncertain due to uncertainties in the UHECR
confinement time, in the planetoid impact rate, and in the number of
magnetized neutron stars.

We thank Attilio Ferrari, Roger Hildebrand, Don Lamb, Angela Olinto,
Simon Swordy, Pasquale Blasi, Willy Benz, Sterling Colgate, Walter
Drugan, Carlo Graziani, Cole Miller, Don Rej and Eli Waxman for
enlightening discussions.  This research was supported by the Center
for Astrophysical Thermonuclear Flashes at the University of Chicago
under Department of Energy contract B341495.

\end{multicols}

\end{document}